\documentstyle[preprint,aps,tighten]{revtex}

\begin{document}

\title{Semiclassical limit of universal parametric density 
correlations}

\author{Alfredo M. Ozorio de Almeida}

\address{Centro Brasileiro de Pesquisas F\'{\i}sicas \\
	 R. Xavier Sigaud, 150, CEP 22290-180  Rio de Janeiro, Brazil}

\author{Caio H. Lewenkopf}

\address{Instituto de F\'{\i}sica, 
	 Universidade do Estado do Rio de Janeiro \\
	 R. S\~ao Francisco Xavier, 524, CEP 20559-900 Rio de Janeiro, 
	 Brazil}

\author{Eduardo R.  Mucciolo}

\address{Departamento de F\'{\i}sica, 
	Pontif\'{\i}cia Universidade Cat\'olica \\
	R. Marqu\^es de S\~ao Vicente, 225, 
	CEP 22453-970 Rio de Janeiro, Brazil }

\date{\today}

\maketitle

\begin{abstract}
Reviewing the semiclassical theory for the parametric level density
fluctuations, we show that for large parametric changes the density
correlation function, after rescaling, becomes universal and coincides
with the leading asymptotic term obtained from Random Matrix
Theory. The advantage of the semiclassical approach is to provide a
simple recipe for the calculation of the non-universal scaling
parameter from elements of the underlying classical dynamics specific
to the system. We discuss recent improvements of the theory, which
introduce a requantization of the smoothed level density.
\end{abstract}

\draft

\narrowtext

\section{Motivation}
\label{sec:I}

One of the interesting new developments in the study of quantum
manifestations of classical chaos concerns parametric
correlations. For over a decade, most of the works on quantum chaos
concentrated in spectral fluctuations of a fixed Hamiltonian. This
large effort led to a solid body of evidence that if a system is
chaotic in the classical limit, its quantum spectrum exhibits
universal fluctuations which coincide with those of a suitable
ensemble of random matrices \cite{LesHouches}. There was also some
parallel activity in trying to understand the statistical behavior of
quantities related to the response of a chaotic Hamiltonian to
external perturbations, such as Landau-Zener transitions
\cite{Wilkinson88} and energy level curvature \cite{Gaspard90}. But it
was only a few years ago that Szafer, Simons, and Altshuler
\cite{Szafer93,Simons93a} showed that a system whose spectrum follows
closely the universal fluctuations predicted by random matrix theory
(RMT) will also present universal {\it parametric} behavior. More
precisely, these authors have concluded that when a chaotic
Hamiltonian depends on some external parameter $X$, any correlation
function of spectral fluctuations taken at different values of $X$
becomes system-independent after a proper rescaling.

In this paper we study some universal spectral features of a quantum
mechanical problem whose Hamiltonian $H$ generates chaotic motion in
its classical limit. Considering a parametric dependence of $H$ on
$X$, the Schr\"odinger equation reads
\begin{equation}
H(X) \Psi_\nu({\bf r};X) = E_\nu(X) \Psi_\nu({\bf r};X) \;.
\end{equation}
In what follows, the parameter $X$ will be associated with any tunable
physical quantity, provided that its variation neither alters
significantly the classical dynamics of the system, nor breaks or
restores any symmetry. The first condition is easily realized in the
semiclassical limit, since classically small parameter variations are
responsible for significant level fluctuations. The second condition
is not difficult to be satisfied in most cases of practical relevance.

Previous works have concentrated on the level velocity correlation
function \cite{Szafer93,Simons93b}
\begin{equation}
C(X) = \frac{1}{\Delta^2} \left\langle \frac{\partial E_\nu }{\partial
\overline{X}}\!\left(\overline{X} - \frac{X}{2}\right) \,
\frac{\partial E_\nu }{\partial \overline{X}}\!\left(\overline{X} +
\frac{X}{2}\right) \right\rangle_{\overline{X},\nu}, \end{equation}
where $\Delta$ denotes the mean level spacing around the state
$\nu$. This function, although straightforward to evaluate
numerically, is not suitable for any exact or systematic analytical
evaluation.  We would like to stress that the difficulty is intrinsic
to the function, since it requires the precise knowledge of every
level as a function of $X$. Therefore, instead of the usual 2-point
Green's function, one would need a combination of arbitrarily high
$N$-point Green's functions, which is very impractical. A few years
ago, Berry and Keating \cite{Berry94} proposed a semiclassical
approximation based on the density correlation function for
$C(X)$. Their result agrees very nicely with the numerical results
\cite{Bruus96} for the large $X$ limit, indicating that the 2-point
Green's function gives a good asymptotic expansion. For the small $X$
region, their analytical expression gives, as one would expect, very
poor results.

For these reasons, in order to study parametric statistics
semiclassically, we will rather focus on a formally more amenable
quantity, namely, the density correlation function
\begin{equation}
K(\Omega,X) = \left\langle
\rho^{fl}\!\left(E-\frac{\Omega}{2},\overline{X} -
\frac{X}{2}\right)\, \rho^{fl}\!\left(E+\frac{\Omega}{2},\overline{X}
+ \frac{X}{2}\right) \right\rangle_{E,\overline{X}},
\end{equation}
where the fluctuating density $\rho^{fl} = \rho - \langle \rho
\rangle$ is defined in terms of the level density
\begin{equation}
\rho(E,X) = -\frac{1}{\pi}\, \mbox {Im} \sum_\nu \frac{1}{E - E_\nu(X)
+ i0^{+}} \;,
\end{equation}
and the average density $\langle \rho \rangle$, or inverse mean level
spacing $1/\Delta$. Sometimes, in the semiclassical approach, it is
convenient to use the cumulative level density, $N(E)$ defined as
$N(E) = \int_{-\infty}^E dE^\prime \rho(E^\prime)$, instead of the
density itself as we shall see in Section IV.

All statistical measures of spectral fluctuations presuppose a constant
average level spacing, which is not guaranteed for general families of
Hamiltonians. Indeed, the mean level spacing for smooth Hamiltonians
is accurately approximated by the Weyl rule: $N^{\mbox{\scriptsize
Weyl}} (E, X) \approx V(E, X)/(2\pi\hbar)^d$, where $d$ is the number
of degrees of freedom and
\begin{equation}
V(E, X) = \int \!d^dp\ d^d q \,\Theta \Big( E - H({\bf p}, {\bf q}, X)
\Big) ~,
\end{equation}
$\Theta$ being the unit step function. It is therefore desirable to
work with the ``volume spectrum" rather than directly with the energy
spectrum, as proposed by Goldberg and collaborators
\cite{Goldberg91}. So as to avoid some technicalities in the classical
theory, we will adopt the alternative procedure of defining an
``unfolded classical Hamiltonian"
\begin{equation}
\label{Hprime}
H^\prime({\bf p}, {\bf q}, X) = V \Big( H({\bf p}, {\bf q}, X), X
\Big) ~.
\end{equation}
The classical motion is then identical to that of the original
Hamiltonian, except for a rescaling in time. Quantizing (\ref{Hprime})
we obtain levels $E_\nu^\prime$ that fluctuate with $X$ about stable
mean values $\nu(2\pi \hbar)^d$, i.e., for the unfolded system,
$\Delta = (2\pi\hbar)^d$. (The primes will be omitted in what
follows.)

The paper is organized as follows: In Section II, we present a
reminder of the main parametric random matrix results, extracting the
relevant asymptotic correlation functions. In Section III the standard
semiclassic theory is discussed, putting emphasis on the diagonal
approximation and its sequels. In Section IV we discuss the
implications for $K(\Omega,X)$ of the recent proposal of Bogomolny and
Keating \cite{Bogomolny96} for requantization of the smoothed level
density. We summarize the results and conclude with a general
discussion about the range of validity of the semiclassical approach
in Section V.

\section{Reminder of the random matrix result}
\label{sec:II}

The random matrix calculations leading to an exact expression for
$K(\Omega,X)$ have become relatively standard. Several references can
serve as introduction to the subject (see, for instance,
\onlinecite{Efetov83,Fyodorov95,Efetov96}). The detailed derivation of
$K(\Omega,X)$ was first given in Ref. \onlinecite{Simons93b} in the
context of disordered metallic systems. The technique employed for
such continuous systems can be easily adapted to discrete
Hamiltonians. It consists in expressing the one-point Green's function
$G_{jl}(E) = \langle l |(E - H + i0^{+})^{-1}| j \rangle$, where $l$
and $j$ label arbitrary basis states, in terms of an integral over $N$
commuting and $N$ anticommuting variables, where $N$ represents the
number of basis states, and then directly performing the average over
a certain Gaussian ensemble of Hamiltonians, $\{H\}_{N\times N}$. The
quartic term thus generated can be decoupled through a
Hubbard-Stratonovich transformation at the expense of introducing a
graded matrix, also known as a {\it supermatrix}. The size and
symmetry of this supermatrix depends on the symmetry class of the
Hamiltonian and the order of the correlation function in question. For
instance, for the Gaussian unitary ensemble (GUE) the calculation of
the density correlation function requires an $8\times 8$ supermatrix
transforming according to the group U(1,1$|$2). After integrating out
the commuting and anti-commuting variables, one then has to carry out
an integration of an effective action that involves the supermatrix
elements. This is usually done in the saddle-point approximation,
exploring the condition $N\gg1$. Notice that this technique greatly
simplifies the problem: The $2N$ variables are reduced to a small,
workable set. However, we should remark that the apparently
straightforward steps just described involve rather difficult
mathematical considerations. For the important details the reader
should consult the references cited above.

One of the most important conclusions drawn from
Refs. \onlinecite{Simons93a,Simons93b} is the existence of a single
quantity controlling the scale of parametric fluctuations of all
spectral functions of chaotic Hamiltonians. This result is obtained
explicitly in the supersymmetric formulation of the disordered
metallic system in the diffusive regime (the so-called zero mode), as
well as (but less surprisingly) in RMT when $N\rightarrow\infty$. More
precisely, if $X$ is the external parameter, the rescaling $x =
X/X_c$, where
\begin{equation}
X_c = \left\langle \left[ \frac{1}{\Delta} \frac{dE_\nu(X)}{dX} 
\right]^2 \right\rangle^{-1/2},
\label{eq:factor}
\end{equation}
eliminates all system dependence in the parametric correlation
functions, up to a global prefactor. We will later argue that a
similar conclusion can be obtained from a purely semiclassical
approach.

We now present the results of Ref.\onlinecite{Simons93b} in the
rescaled form, assuming also that all energies are expressed in terms
of $\Delta$, introducing $\omega = \Omega/\Delta$. In the absence of
time-reversal symmetry, the final result for the density correlation
function is
\begin{equation}
\label{eq:k_GUE}
k^{GUE} (\omega,x) = \frac{1}{2}\,\mbox{Re} \!\int^1_{-1} d\lambda
\int_1^{\infty} d\lambda_1 \, \exp \left[-F^{GUE} (\lambda,\lambda_1)
\right]~,
\end{equation}
where $k=K/\langle \rho \rangle^2$ and the free energy is given by
\begin{equation}
F^{GUE} = \frac{\pi^2x^2}{2} (\lambda_1^2 - \lambda^2) +
i\pi\omega^{+} (\lambda - \lambda_1) ~,
\label{eq:F_GUE}
\end{equation}
with $\omega^+ = \omega + i0^+$. The integration in (\ref{eq:k_GUE})
can be carried out explicitly, and gives
\begin{equation}
\label{eq:k_GUEex}
k^{GUE}(\omega,x) = \frac{1}{4x^2}\,\mbox{Im} \left\{
\mbox{erfc}\left(\frac{\pi x + i \omega/x}{\sqrt{2}}\right)
\left[ \mbox{erfc}\left(\frac{-\omega/x+i\pi x}{\sqrt{2}}\right) -
       \mbox{erfc}\left(\frac{-\omega/x-i\pi x}{\sqrt{2}}\right)
\right] \right\} ~.
\end{equation}
where erfc is the complementary error function \cite{Abramowitz}.

For systems where time-reversal symmetry is preserved, modelled by the
Gaussian Orthogonal Ensemble (GOE), the density correlation function
is
\begin{equation}
\label{eq:k_GOE}
k^{GOE}(\omega,x) = \mbox{Re}\int^1_{-1}\! d\lambda \int_1^{\infty}\!
d\lambda_1 \int_1^\infty \!d\lambda_2 \, \frac{(1-\lambda^2) (\lambda -
\lambda_1 \lambda_2)^2 \exp \left[ -
F^{GOE}(\lambda,\lambda_1,\lambda_2) \right]}{(2 \lambda \lambda_1
\lambda_2 - \lambda^2 - \lambda_1^2 - \lambda_2^2 + 1)^2}\;,
\end{equation}
with the free energy 
\begin{equation}
F^{GOE} = \frac{\pi^2x^2}{4} (2 \lambda^2_1 \lambda_2^2 - \lambda_1^2
- \lambda_2^2 - \lambda^2 + 1) + i\pi\omega^{+} (\lambda - \lambda_1
\lambda_2) ~.
\label{eq:F_GOE}
\end{equation}
In distinction to the previous case, here the triple integral in
(\ref{eq:k_GOE}) cannot be expressed in a closed form.

We will now focus on a regime when one either has $x\gg 1$ or
$\omega\gg 1$. Considering initially large variations of the external
parameter, one finds that the leading contributions to the multiple
integrals come from the regions around two points: $\lambda =
\lambda_1 = \lambda_2 = 1$ (A) and $\lambda = -\lambda_1 = -\lambda_2
= -1$ (B). In passing, we remark that there exists a mapping of RMT
into the problem of many particles interacting by a $1/r^2$ potential
in one dimension \cite{1Danalog}. In that context, the contribution
(A) gives the ``hydrodynamical'', $q=0$ limit of the particle density
correlation function, whereas (B) is related to the ``Friedel'',
$q=2k_F$ oscillations. Here, (A) is related to the monotonic,
non-oscillating decay of the correlation function; (B), on the other
hand, yields a series of oscillating terms in $\omega$. To obtain an
asymptotic expression for $k(\omega,x)$, we expand the free energies
around these points and retain the (linear in $\lambda$'s)
lowest-order terms. Performing the exponential integrals (see the
Appendix \ref{appendA} for details), we find that
\begin{equation}
k^{GUE}(\omega,x) \approx 
 -\frac{1}{2}\,
  \frac{\pi^2\omega^2 - \pi^4 x^4}{(\pi^2\omega^2 + \pi^4 x^4)^2} 
 +\frac{1}{2}\,
  \frac{\cos(2\pi \omega)}{\pi^2\omega^2 + \pi^4x^4}
\label{asympdgue}
\end{equation}
and
\begin{equation}
k^{GOE}(\omega,x) \approx - \frac{\pi^2\omega^2 - \pi^4 x^4
/4}{(\pi^2\omega^2 + \pi^4 x^4 /4)^2} + \frac{1}{2}
\frac{\cos(2\pi\omega)} {(\pi^2\omega^2 + \pi^4 x^4 /4 )^2} ~.
\label{asympdgoe}
\end{equation}
The right-hand sides of both equations represent the leading terms of
an asymptotic expansion on both $x$ and $\omega$ whenever $|\pi^2
x^2\beta/2 - i\pi\omega|\gg 1$, with $\beta=1$ for the GOE and
$\beta=2$ for the GUE. At $x=0$ one has the usual $1/\omega^2$ decay
of the density fluctuations, plus the first oscillating terms. Notice
also that the oscillating and non-oscillating terms are held only to
their lowest order in $x^{-2}$ and $\omega^{-2}$. For the GOE, this
means that we are not showing a non-oscillating term of order
$\omega^{-4}$. For the GUE, higher orders terms can be obtained
without much effort by recalling (\ref{eq:k_GUEex}).

The energy oscillations present in Eqs. (\ref{asympdgue}) and
(\ref{asympdgoe}) are related to the quantum (discrete) nature of the
energy spectrum. For the GUE case they appear even at the lowest order
in $\omega^{-2}$; in fact, for $x=0$, the asymptotic form
(\ref{asympdgue}) is actually exact and applies to all $\omega$
scales. For the GOE case, since level repulsion is less pronounced,
the oscillations only show up at the next-lowest order, namely, they
are weaker corrections to the leading $\omega^{-2}$ decay.

One is tempted to imagine that the $\omega$ oscillations should not be
present in a formulation where there is no intrinsic energy level
quantization. Indeed, in the following section we will verify that the
semiclassical trace formula (applied in the usual way and based on the
assumption that the underlying classical dynamics is chaotic) only
yields the monotonic decay of $k(\omega,x)$. Still, a problem emerges
as to whether one can systematically obtain oscillating terms, or any
parametric correlation function beyond the asymptotic limit, in a
semiclassical framework. This question was partially answered by
Bogomolny and Keating \cite{Bogomolny96} when they calculated
$k(\omega,x=0)$ using an expansion in periodic orbits and forced
energy level quantization as an additional constraint. We will get
back to this point later.

Another important aspect, often neglected, but particularly relevant
to mesoscopic phenomena, is the applicability range of the
semiclassical approach used in the study of parametric variations. For
this purpose, the calculations below will be carried out with special
care.

\section{Semiclassical approach}
\label{sec:III}

The fluctuating part of the level density $\rho^{fl}(E,X)$ can be
smoothed over an energy range $\eta/\hbar$ and expressed, in the
semiclassic limit, by the Gutzwiller trace formula \cite{Gutzwiller90}
\begin{equation}
\label{rhoGut}
\rho^{fl}(E,X) = \frac{1}{2\pi \hbar}\,
 \sum_{\gamma,r} \frac{T_\gamma}{|\det
(M^r_\gamma - I)|^{1/2}} \exp \left( \frac{i}{\hbar} r S_\gamma -
\frac{i\pi}{2} r \nu_\gamma -\frac{\eta}{\hbar} |r| T_\gamma\right) 
 ~,
\end{equation}
where $T_\gamma$, $M_\gamma$, $S_\gamma$, and $\nu_\gamma$ stand,
respectively, for the period, monodromy matrix, action, and Maslov
index of a periodic orbit labelled by $\gamma$. The sum is performed
over all primitive orbits $\gamma$ and their repetitions $r$ (positive
and negative), irrespective of multiplicity due to time-reversal or
other symmetries. The smoothing of $\rho^{fl}$ can be done in other
ways than the exponential cutoff introduced in (\ref{rhoGut}). Later
on, we shall also cutoff the sum over periodic orbits sharply at a
certain period $T^*$. Both $h/T^*$ and $\eta/2\pi$ are always taken
larger or equal to $\Delta$.

Let us write the correlation function $K(\Omega,X)$ in terms of the
semiclassical level density as given by Eq.~(\ref{rhoGut}),
\begin{eqnarray}
K^{sc}(\Omega,X) & = & \frac{1}{(2\pi \hbar)^2} \Bigg\langle
\sum_{{\gamma,r}, {\gamma^\prime,r^\prime}} A_{\gamma r}\! \left (E +
\frac{\Omega}{2}, \overline{X} + \frac{X}{2} \right) A_{\gamma^\prime
r^\prime}\!\left( E - \frac{\Omega}{2}, \overline{X} - \frac{X}{2}
\right) \nonumber \\ & & \times \exp \left\{ \frac{i}{\hbar} \left[ r
S_\gamma \!\left( E + \frac{\Omega}{2},\overline{X}+\frac{X}{2} \right)
- r^\prime S_{\gamma^\prime}\! \left( E -
\frac{\Omega}{2},\overline{X}-\frac{X}{2} \right) \right] \right.
\nonumber\\ & & \left. ~~~~~~~~~~~ + i \frac{\pi}{2} (r \nu_\gamma -
r^\prime \nu_{\gamma^\prime}) - \frac{\eta}{\hbar} ( |r| T_\gamma +
|r^\prime| T_{\gamma^\prime} ) \right\} \Bigg\rangle_{E,\overline{X}},
\end{eqnarray}
where $A_{\gamma r} = T_\gamma |\det (M_\gamma^r - I)|^{-1/2}$.
In order to further proceed analytically, we restrict the calculations
to a regime where classical perturbation theory is applicable. This
means that we shall consider variations of $E$ and $X$ which
classically imply a small change in the actions $S_\gamma$. Such a
change can nonetheless be very large in the scale of $\hbar$,
corresponding to large quantum effects. Under this condition, it is a
good approximation to write
\begin{eqnarray}
\label{Sexpansion}
S_\gamma \left( E \pm \frac{\Omega}{2},\overline{X} \pm \frac{X}{2}
\right) &=& S_\gamma (E,\overline{X}) \pm \frac{\partial
S_\gamma}{\partial E} \,\frac{\Omega}{2} \pm \frac{\partial
S_\gamma}{\partial \overline{X}} \, \frac{X}{2} + \cdots \nonumber \\
&\equiv & S_\gamma (E,\overline{X}) \pm T_\gamma (E,\overline{X})
\,\frac{\Omega}{2} \pm Q_\gamma (E,\overline{X}) \,\frac{X}{2} +
\cdots ~,
\end{eqnarray}
which defines $Q_\gamma$ as the parametric velocity of the orbit
action. The general expression for the parametric variation of the
action is presented in Appendix B. In the absence of any dependence on
Planck's constant, one can neglect the energy (or parameter)
corrections to $A_\gamma$ as compared to the strong energy dependence
of the exponential term. In this approximation, consider the
evaluation of
\begin{equation}
\label{averageoverE}
I_{\gamma \gamma^\prime}(T) = \left\langle \mbox{e}^{i/\hbar\left[
       S_\gamma (E,\overline{X}) - S_{\gamma^\prime} (E,\overline{X}) 
       \right]}\right\rangle_E ~.
\end{equation}
For orbits with period shorter than the Heisenberg time $t_H \equiv
h/\Delta$, one does not in general expect to find pairs of orbits
$(\gamma,\gamma^\prime)$ with actions differing by less than $\hbar$,
unless they are symmetry related orbits. As a consequence, upon energy
averaging, $I_{\gamma \gamma^\prime}(T<t_H) \approx
g_\gamma\delta_{\gamma,\gamma^\prime}$, where $g_\gamma$ is the
multiplicity of symmetry-related orbits. (We take here the simplest
case where any symmetry holds for all considered parameter values.)
This is the essence of the diagonal approximation. On the other hand,
due to the exponential proliferation of orbits, for sufficiently long
times, i.e., for a small smoothing parameter $\eta$, one enters a
regime where there is an abundance of pairs of periodic orbits
$(\gamma,\gamma^\prime)$ satisfying $S_\gamma - S_{\gamma^\prime} <
\hbar$. This situation is expected to occur for time intervals longer
than $t_H$ (or, correspondingly, for $E<\Delta$). Under these
circumstances, the diagonal approximation is no longer valid.
Actually, for such long times even the validity of the trace formula
is problematic, so that the unsmoothed energy spectrum should be
resummed \cite{Berry92}. At this point the necessity of smoothing
becomes evident: it keeps the approximations under control.

Restricting ourselves to the range where the diagonal approximation is
valid, we have
\begin{eqnarray}
\label{eq:16}
K^{sc}_D (\Omega,X) & = & \frac{1}{(2\pi \hbar)^2}\, \left\langle
\sum_{\gamma,r} g_\gamma |A_{\gamma r}(E,\overline{X})|^2 \right.
\nonumber \\ & & \left.  \times \exp\!\left\{ \frac{i}{\hbar} \left[ r
T_\gamma(E, \overline{X})\,\Omega + r Q_\gamma(E,\overline{X})\, X
\right] - 2 \frac{\eta}{\hbar} |r| T_\gamma \right\}\right
\rangle_{E,\overline{X}} ~.
\end{eqnarray}
To evaluate the average in (\ref{eq:16}), we first notice that, in a
fully chaotic regime, the proportion of high period orbits that are
repetitions of lower periods is exponentially small, so we neglect all
$|r|\neq 1$. In the next step we recast (\ref{eq:16}) as an integral
by defining the smooth interpolating functions
${A^2}(E,\overline{X},t)$ through the relation
\begin{equation}
\int \! dt {A^2}(E,\overline{X},t) \sum_\gamma \delta (t - T_\gamma) =
\sum_\gamma A^2_\gamma (E, \overline{X}) ~.
\end{equation}
Further simplification can be achieved using the uniformity principle
over periodic orbits (also known as the Hannay-Ozorio de Almeida sum
rule) \cite{Hannay84}
\begin{equation}
\label{sumrule}
{A^2}(E,\overline{X},t) \sum_\gamma \delta (t -
T_\gamma)~~\begin{array}{c} \\
\relbar\joinrel\relbar\joinrel\relbar\joinrel\longrightarrow \\
\mbox{\scriptsize{$|t|\rightarrow \infty$}}\end{array}~~ |t| \;.
\end{equation}
It is important to notice that this result independs on the
``center-of-mass'' variables $E$ and $\overline{X}$. We can then
evaluate
\begin{equation}
K^{sc}_D (\Omega,X) = \frac{g}{2(\pi\hbar)^2} \mbox{Re}
\int_{\tau}^\infty \! dt \,|t| \, \mbox{e}^{it\Omega/\hbar - 2\eta
|t|/\hbar} \left\langle \mbox{e}^{iQ(E,\overline{X},t)X/\hbar}
\right\rangle.
\label{eq:newalf}
\end{equation}
The lower limit $\tau$ is the period of the shortest periodic
orbit. Of course it is simplistic to extrapolate the uniformity
principle this far, but it is easy to improve the theory by including
a few short orbits individually and hence increasing $\tau$.

The average in Eq. (\ref{eq:newalf}) is evaluated over periodic orbits
with fixed period $t$. In view of the fact that the rest of the
integrand oscillates rapidly in time, it is important that the finite
range in $E$ and $\overline{X}$ allows many orbits with a given large
period. There being two parameter families of periodic orbits, one
should take $Q_\gamma(E(\overline{X}),\overline{X},t)$ and
subsequently average only over $\overline{X}$, or vice-versa. For the
energy correlation with no variation of parameters, there will be
isolated periodic orbits of given period in the energy range where the
averaging is carried out. Thus, for fixed $\overline{X}$, the average
in Eq. (\ref{eq:newalf}) runs over a large number of periodic orbits
with parametric velocity $Q_\gamma(t)$, which may be considered as
samples of the probability distribution $P_t(Q)$. The latter is
assumed to be Gaussian,
\begin{equation}
P_t(Q) = \frac{1}{\sqrt{2\pi \overline{Q^2}(t)}} \exp \left[
\frac{-Q^2}{2\overline{Q^2}(t)} \right],
\label{eq:gaussantz}
\end{equation}
because our theory involves arbitrary parametric variations
$H(X)$. Even if there exist correlations among the actions of the
periodic orbits $\gamma(t)$, as postulated by \cite{Argaman93}, these
will in general be broken by the parametric velocities
\begin{equation}
\label{defQ}
Q_\gamma =  \int_0^{T_\gamma} \! dt \,\frac{\partial H}{\partial X} 
	\Big({\bf p}(t), {\bf q}(t), X\Big) \;,
\end{equation}
evaluated along each orbit. The Gaussian width is a function of
$\overline{X}$ and $\overline{E}$ (the center of the energy range),
being defined as
\begin{equation}
\overline{Q^2}(t) = \frac{1}{N(t)} \sum_{\gamma(t)} Q^2_{\gamma(t)},
\end{equation}
where $N(t)$ is the number of periodic orbits with period $t$. Note
that we take $\overline{Q}=0$, since the shape of the shell for the
unfolded system is affected by $X$, but not its volume, so $\partial
H/\partial X$ averaged over the shell is zero. Although the Gaussian
assumption is not quite at a par with the uniformity principle, the
arguments presented above make it plausible. Nothing is yet known of
the relation between higher moments $\overline{Q^n}$ and
$\overline{Q^2}$, so as to obtain a more rigorous result. Numerical
simulations for the particular case where the external parameter is a
magnetic flux line \cite{Bruus96} provide further support to
(\ref{eq:gaussantz}).

The average in Eq. (\ref{eq:newalf}) can be evaluated using
(\ref{eq:gaussantz}):
\begin{equation}
\left\langle e^{i QX/\hbar} \right\rangle = \exp \left[
-\frac{X^2}{2\hbar^2} \overline{Q^2}(\overline{E},\overline{X},t)
\right].
\label{eq:newaveralf}
\end{equation}
As we commented previously, the energy average was already utilized in
obtaining the periodic orbits with fixed period. Strictly, we should
now average (\ref{eq:newaveralf}) over $\overline{X}$, but both
$\overline{Q^2}$ and the Gaussian (\ref{eq:gaussantz}) are smooth
functions of $\overline{X}$, so we may just evaluate the latter at the
center of the averaging range. As for the time dependence of the mean
square parametric velocity, we refer to previous treatments
\cite{Goldberg91,Bohigas95}, leading to the result
\begin{equation}
\label{Q2t}
\overline{Q^2} = \alpha |t|, \qquad \mbox{with} \qquad
\alpha(\overline{E}, \overline{X}) = 2\int_0^\infty dt \left\langle
\frac{\partial H}{\partial X} \Big( p(t),q(t) \Big) \frac{\partial
H}{\partial X} \Big( p(0),q(0) \Big) \right\rangle_{e.s.} ~.
\end{equation} 
Here, the average over the periodic orbits has been substituted by an
average over the entire energy shell, in accordance with the
uniformity principle. It is important to note that the decay of the
classical correlation function in (\ref{Q2t}) need only be integrable,
since many available chaotic systems do not exhibit full exponential
decay of the correlations. Strictly, the linear dependence of
$\overline{Q^2}$ on $t$ should only hold for times longer than that of
the decay of the correlation function, but numerical investigations
\cite{Bruus96,Pluhar95} have found this feature to be quite robust.
(Notice that there is a factor 2 missing in the definition of $\alpha$
in Ref.~\onlinecite{Bruus96}. Modifying $\alpha$ accordingly the
semiclassical prediction for $X_c$ changes by a factor $\sqrt{2}$. As
a result, the agreement between the numerics and the semiclassical
theory becomes excelent.)

Substituting (\ref{Q2t}) and (\ref{eq:newaveralf}) into
Eq. (\ref{eq:newalf}) and evaluating the integral in the limit
$\tau\rightarrow 0$, we finally arrive at the main result of this
section
\begin{equation}
\label{Kstandard}
K^{sc}_D (\Omega,X) = - \frac{1}{\beta\pi^2} \, \frac{\Omega^2 -
\left(\alpha X^2/2\hbar + 2\eta\right)^2} {\left[\Omega^2 +
\left(\alpha X^2/2\hbar + 2\eta\right)^2 \right]^2} ~,
\end{equation}
where $\beta=2/g$. The nongeneric long-wave oscillations in the
correlation, characteristic of each individual system, appear as we
choose $\tau$ equal or greater to the period of the shortest periodic
orbit. The above expression corresponds to the leading non-oscillatory
term of the asymptotic expansion of the correlation function exactly
calculated within RMT [Eqs. (\ref{asympdgue}) and (\ref{asympdgoe})],
and hence presumed to hold for generic chaotic systems.

On the other hand, this formalism gives a recipe for calculating
$\alpha$, which is a system-specific quantity and fixes the value of
the parametric rescaling factor $X_c$. From this point-of-view, the
semiclassical method and the RMT are complementary for a full
understanding of $K(\Omega, X)$.  An important immediate application
concerns possible transitions from GOE families to GUE families of
Hamiltonians. If this occurs for a small variation of a second
parameter as in \cite{Bohigas95}, the classical parameter $\alpha$
should remain unchanged, whereas $X_c$ must grow because of increased
level repulsion. By comparing the invariant form of
Eq. (\ref{Kstandard}) with (\ref{asympdgue}) and (\ref{asympdgoe}), we
can predict that $X_c$ will increase by a factor $\sqrt{2}$.

\section{Requantization and its implications for the density correlation 
function}
\label{sec:IV}

In a recent paper, Bogomolny and Keating \cite{Bogomolny96} proposed a
new semiclassical procedure to obtain the density correlation function
$K(\Omega, 0)$ and its parametric extension, $K(\Omega, X)$. Their
result for $K(\Omega, 0)$ coincides with the exact GUE expression for
systems with broken time-reversal invariance and gives the leading GOE
oscillatory and non-oscillatory terms in powers of $\omega^{-2}$ for
time-reversal symmetric systems. We will here rederive their results
within the wider context of parametric variations.

The starting point of the derivation in Ref. \onlinecite{Bogomolny96}
is standard and considers the sum over periodic orbits in the
Gutzwiller trace formula (\ref{rhoGut}) up to some value of $T^*$. As
a consequence, the information about energy scales smaller than
$\hbar/T^*$ is washed out, and strictly speaking, the connection to
RMT results cannot be made in such an energy range. However, one can
still introduce information related to the discrete nature of the
spectrum through the following ``requantization'' condition
\cite{Aurich92}
\begin{equation}
N_{T^*}\Bigl(E_\nu(T^*)\Bigr) = \nu + \frac{1}{2} \;,
\end{equation}
by which, for a given value of $T^*$, one has a well-defined procedure
to obtain a set of eigenvalues $\{ E_\nu(T^*)\}$. It has to be
stressed that such a procedure does not imply any control on
accuracy. This is equivalent to say that, by increasing the value of
$T^*$, it is not guaranteed that $\{ E_\nu(T^*)\}$ will converge to
the exact spectrum. Actually, this quantization rule is {\sl not}
applicable for $T^* > t_H$ since for such periods the cumulative
semiclassical level density can become a non-monotonic function.

The requantized level density, defined as
\begin{equation}
D_{T^*}(E,X) = \sum_\nu \delta \Bigl( E - E_\nu(T^*, X)\Bigr),
\end{equation}
can be rewritten in the form
\begin{eqnarray}
D_{T^*} (E,X) = &&\rho_{T^*} (E,X)\sum_\nu \delta \Bigl(N_{T^*}(E,X) -
\nu - 1/2 \Bigr) \nonumber \\ = & & \rho_{T^*}(E,X)
\sum_{k=-\infty}^{+\infty} (-1)^k \exp\left[ 2\pi i k
N_{T^*}(E,X)\right],
\label{eq:DT}
\end{eqnarray}
with $\rho_{T^*}(E,X)$ standing for the parametric level density given
by the truncated trace formula. It is immediate to see that averaging
Eq. (\ref{eq:DT}) yields
\begin{equation}
\big\langle D_{T^*}(E,X) \big\rangle_{E,X} = 
	 \big\langle \rho_{T^*}(E,X)\big\rangle_{E,X} 
	 \equiv \rho^{\mbox{\scriptsize Weyl}}(E, X) 
\;,
\end{equation}
equivalent to the usual level density, easily obtained from the Weyl 
formula.

The advantages of this procedure become evident once we analyze the
spectral fluctuations. Let us study the 2-point spectral correlation
function for $D_{T^*} (E)$ in order to compare with the results of the
preceding sections. It is useful to write the density correlation
function in the form
\begin{equation}
K(\Omega, X) = \left\langle 
   D_{T^*}\left(E + \frac{\Omega}{2}, \overline{X} + \frac{X}{2}\right)
   D_{T^*}\left(E - \frac{\Omega}{2}, \overline{X} - \frac{X}{2}\right)
	    \right\rangle_{E,\overline{X}} - \Big(\rho^{\mbox{\scriptsize
	    Weyl}}\Big)^2 ~.
\end{equation}
The averages are defined as in Section III. Thus,
\begin{eqnarray}
\label{R2new}
& & \!\!\!\!\!\!\!\!\!\!\!\!\!\!\!\!
   K(\Omega,X) = \left\langle \rho_{T^*}\left(E + \frac{\Omega}{2},
   \overline{X} + \frac{X}{2}\right) \rho_{T^*}\left(E -
   \frac{\Omega}{2}, \overline{X} - \frac{X}{2}\right)
   \right. \nonumber \\ ~~~~& \times & \left. \!\!\sum_{k_1, k_2}
   (-1)^{k_1 - k_2} \exp\! \Bigg( 2\pi i \left[ k_1 N_{T^*}\left(E +
   \frac{\Omega}{2}, \overline{X}+ \frac{X}{2}\right) - k_2
   N_{T^*}\left(E - \frac{\Omega}{2}, \overline{X}- \frac{X}{2}\right)
   \right]\Bigg) \right\rangle_{E,\overline{X}} \nonumber \\ & & -
   \Bigl(\rho^{\mbox{\scriptsize Weyl}} \Bigr)^2.
\end{eqnarray}
One can recover the standard semiclassical result discussed in the
previous section by considering the simplest term in the double sum,
namely, $k = k_1 = k_2 = 0$:
\begin{equation}
\label{Kstand}
K^{k=0}(\Omega, X) = \left\langle
\rho_{T^*}^{fl}\left(E+\frac{\Omega}{2},\overline{X}+\frac{X}{2}\right)
\rho_{T^*}^{fl}\left(E-\frac{\Omega}{2},\overline{X}-\frac{X}{2}\right)
\right\rangle_{E,\overline{X}} \;.
\end{equation}
This is the same calculation as presented in the previous section,
except that now the exponential cutoff is substituted by a sharp
truncation at $T^*$. This causes a spurious oscillatory term which
only becomes negligible in the asymptotic regime where $\alpha
X^2/2\hbar \gg \Delta$. In other words, the semiclassical result only
becomes independent of the cutoff at the asymptotic limit, where it
coincides with RMT. In Ref.~\onlinecite{Bogomolny96}, this term with
$k_1=k_2=0$ was called ``diagonal approximation".  This should not be
confused with the standard diagonal approximation discussed in the
previous section.

We are left to discuss the terms with $k_1 \neq 0$ and $k_2 \neq 0$.
It is easy to see that the ones with $k_1 \neq k_2$ give a negligible
contribution to the correlation function. Indeed, dividing the
spectrum for the unfolded system (\ref{Hprime}) into a smooth and a
fluctuating part $(N_{T^*} = N^{Weyl} + N^{fl}_{T^*})$ we have
$N^{Weyl}(E,X)= E/(2\pi\hbar)^d$. For this part, the average
(\ref{R2new}) furnishes
\begin{eqnarray}
 \!\!\!\!\!\!&&\left| \left\langle \exp\left\{ 2\pi i \left[ k_1
N^{\mbox{\scriptsize Weyl}} \left( E + \frac{\Omega}{2}, \overline{X}
+ \frac{X}{2} \right) - k_2 N^{\mbox{\scriptsize Weyl}} \left( E -
\frac{\Omega}{2} , \overline{X}-\frac{X}{2} \right) \right] \right\}
\right\rangle_{E,\overline{X}} \right|= \nonumber \\ && ~~~~~~~~~~~~~~~
\frac{(2\pi\hbar)^d}{(k_1-k_2)\delta E} \sin \left[
\frac{(k_1-k_2)\delta E}{(2\pi\hbar)^d} \right]
=\frac{\sin[(k_1-k_2)\delta N]}{(k_1 - k_2) \delta N}\; ,
\end{eqnarray}
which is negligible if the averaging region $\delta E$ extends over
many states, unless $k_1=k_2$. Thus,
\begin{equation}
\label{Ksumk}
K^{k\neq 0}(\Omega, X) = \frac{1}{4\pi^2} \frac{\partial ^2}{\partial
      \Omega_1 \partial \Omega_2} \sum_{k\neq 0} \frac{1}{k^2}
      \exp\left[ 2\pi i k (\Omega_1 - \Omega_2)
      \rho^{\mbox{\scriptsize Weyl}} \right]
      \Phi_k(\Omega_1, \Omega_2, X_1, X_2)
\end{equation}
(evaluated at $\Omega_1=-\Omega_2=\Omega$ and $X_1=-X_2=X$), where $k
\equiv k_1 = k_2$ and
\begin{equation}
\label{phik}
\Phi_k(\Omega_1, \Omega_2, X_1, X_2) = \left\langle \exp \left\{ 2\pi
i k \Big[ N^{fl}_{T^*}(E+\Omega_1, \overline{X}+X_1) -
N^{fl}_{T^*}(E+\Omega_2, \overline{X}+X_2) \Big] \right\}
\right\rangle_{E, \overline{X}} \;.
\end{equation}
One is now left with the formidable task of averaging (\ref{phik}),
for which there is no clear controllable strategy. Bogomolny and
Keating \cite{Bogomolny96} proposed a kind of Gaussian ansatz, which
in practice simplifies the average by making $\langle \exp[i G(E)]
\rangle = \exp[- \langle G^2(E) \rangle/2]$.  Although appealing, this
ansatz is harder to justify classically than a similar passage in the
previous section. Moreover, it is not clear how to implement the
averaging in a consistent way. In Appendix C, we justify the
validity of the Gaussian assumption (for large energy differences
as compared to $\Delta$) using solely the diagonal
approximation Based on that, we have
\begin{eqnarray}
\label{phikgaus}
\Phi_k(\Omega_1, \Omega_2, X_1, X_2) & = & \exp \Bigg\{ -2\pi^2 k^2
 \Bigg\langle \left[ N_{T^*}(E+\Omega_1, \overline{X}+X_1)
 \right. \nonumber \\ & & \left. ~~~~~~~-
 N_{T^*}(E+\Omega_2, \overline{X}+X_2) \right]^2 \Bigg\rangle_{E,
 \overline{X}} \Bigg\} \; .
\end{eqnarray}
There are at least two straightforward ways to evaluate the average in
(\ref{phikgaus}), both giving the same final result. The first one is
by direct subtraction of the cumulative densities, followed by the use
of the diagonal approximation. The second one has the nice property of
putting in evidence some of the delicate points of the semiclassical
averaging procedure, and will be discussed now. An important point to
have in mind, however, is that the average in the exponent of
Eq. (\ref{phikgaus}) is just the usual number variance
$\Sigma^2(\Omega)$ when $X_1=X_2$. If we were interested in $X=0$
correlations at $\Omega\ll\Delta$ and simply substituted into
Eq. (\ref{phikgaus}) the exact RMT expression for $\Sigma^2(\Omega)$,
we would obtain an incorrect result for the level density correlation
function. We interpret this fact as a clear indication that the
Gaussian ansatz of Ref.[10], which neglects correlation between
orbits, is problematic for $\Omega/\Delta \ll 1$. This limitation will
appear more explicitly below.

Let us start writing (\ref{phikgaus}) as
\begin{equation}
\Phi_k(\Omega_1, \Omega_2, X_1, X_2) = \left[ \frac{\Lambda(\Omega_1,
\Omega_2, X_1, X_2)}{\Lambda_0}\right]^{k^2}
\label{eq:theone}
\end{equation}
with
\begin{equation}
\label{Lambdastuff}
\ln \Lambda (\Omega_1, \Omega_2, X_1, X_2) = 4\pi^2 \Big\langle
   N_{T^*}^{fl}(E + \Omega_1, \overline{X} + X_1) N_{T^*}^{fl}(E +
   \Omega_2, \overline{X} + X_2) \Big\rangle_{E,\overline{X}} \;.
\end{equation}
and $\Lambda_0 = \Lambda(0,0,0,0)$. Then, we evaluate 
$\Lambda_0$ using the diagonal approximation (see for
instance \cite{Berry85}),
\begin{equation}
\label{eq:49}
\left \langle \Big[ N_{T^*}(E, \overline{X}) \Big]^2\right \rangle_{E,
\overline{X}} = \frac{g}{2\pi^2} \int_{\tau}^{T^*} \! dt \frac{1}{t} =
\frac{g}{2\pi^2} \ln \left( \frac{T^*}{\tau} \right).
\end{equation}
The lower bound of integration is of the order of the shortest
periodic orbit, beyond which we cannot even extrapolate the uniformity
principle. Recalling that in our case the energy of the shell is
identified with its volume, we obtain, from purely geometrical
considerations, that
\begin{equation}
\tau = \frac{\partial S}{\partial E} \propto E^{-(1-1/d)} \,.
\end{equation}
If we now choose $T^*= \kappa t_H$ ($t_H$ is the Heisenberg time),
Eq. (\ref{eq:49}) becomes
\begin{equation}
\left\langle \left[ N_{\kappa t_H}(E,\bar{X}) \right]^2 \right\rangle
= \frac{2}{g\pi^2} \left( 1 - \frac{1}{d} \right) \ln \left(
\frac{E}{\Delta} \right) + \mbox{constant}\ .
\end{equation}
It is interesting that we recover the RMT result only in the limit
where the phase space dimension $d\rightarrow\infty$. (Incidentaly,
the above formula accounts for the spectral rigidity of the integrable
limit of $d=1$.)

Applying now the diagonal approximation to the numerator of
(\ref{eq:theone}), we write
\begin{equation}
\ln \Lambda(\Omega_1, \Omega_2, X_1, X_2) = 2g\ \mbox{Re}
\int_{\tau}^{T^*} dt \frac{1}{t} \exp \left[ \frac{i}{\hbar}(\Omega_1
- \Omega_2) t - \frac{\alpha(X_1 - X_2)^2 t}{2\hbar^2} \right] ~,
\label{eq:upper}
\end{equation}
which can be expressed in a compact form as
\begin{equation}
\ln \Lambda(\Omega_1, \Omega_2, X_1, X_2) = 2g\ \mbox{Re} \left[
\mbox{E}_1 (\xi \tau) - \mbox{E}_1 (\xi T^*) \right] ~,
\end{equation}
with $\xi = -i(\Omega_1 - \Omega_2)/\hbar + \alpha(X_1 -
X_2)^2/2\hbar^2$ and E$_1$ denoting the exponential integral
\cite{Abramowitz}. Recalling that $T^*$ is of the order of $h/\Delta$
and $\tau$ is a much shorter time scale, we have $|\xi T^*| \gg 1$ and
$|\xi \tau| \ll 1$. As a consequence, E$_1(\xi T^*)$ is exponentially
small and E$_1(\xi \tau)$ can be expanded up to first order
\cite{Abramowitz}. The result is
\begin{eqnarray}
\label{lnLambda}
\ln \Lambda(\Omega_1, \Omega_2, X_1, X_2) &\approx & 2g\ \mbox{Re}
\left[ -\gamma - \ln(\xi \tau) \right] \nonumber\\ & \approx & 2g\
\mbox{Re} \left[ -\gamma + \ln \left( \frac{T^*}{\tau} \right) - \ln
(\xi T^*) \right] ~,
\end{eqnarray}
where $\gamma$ is the Euler constant.
Collecting the above results into (\ref{phikgaus}), we see that the
first two terms in the r.h.s. of (\ref{lnLambda}) are cancelled by
$\ln \Lambda_0$, and
\begin{equation}
\Phi_k(\Omega_1, \Omega_2, X_1, X_2) = \left| \left[
-\frac{i}{\hbar}(\Omega_1 - \Omega_2) + \frac{\alpha(X_1 -
X_2)^2}{2\hbar^2} \right] e^\gamma T^* \right|^{-2gk^2} ~.
\end{equation}
Hence, the sum over $k$ in (\ref{Ksumk}) is an asymptotic series in
inverse powers of $\xi T^*$. To obtain the parametric generalization
of Bogomolny and Keating result \cite{Bogomolny96}, we cutoff this
series in the first term and choose $T^* e^\gamma = t_H/2$, obtaining
\begin{equation}
K^{|k|=1}(\Omega, X) = \frac{\cos(2\pi\Omega/\Delta)}{2 \Delta^2}
      \left| \frac{\Delta}{-i\pi\Omega +
      \alpha\pi^2 X^2/h} \right|^{4/\beta} ~,
\label{eq:final}
\end{equation}
which, after proper rescaling, is exactly (\ref{asympdgue}) or
(\ref{asympdgoe}), depending on the symmetry class labelled by
$\beta=2/g$.

It thus appears that requantization has reproduced precisely the
asymptotic form of the correlation function calculated exactly within
RMT if we postulate the same factor of $\sqrt{2}$ for the change in
$X_c$ from GOE to GUE. However, in taking a sharp cutoff for the
periodic orbit sum, we add the spurious oscillatory term
\begin{equation}
\frac{g T^* \cos \left( \frac{\Omega T^*}{\hbar} - \phi \right) \exp
\left( -\frac{\alpha X^2 T^*}{2\hbar^2} \right)} {2\pi^2 \hbar \left[
\Omega^2 + \left( \frac{\alpha X^2}{\hbar} \right)^2 \right]}
\end{equation}
to (\ref{Kstandard}) with $\eta=0$ [above, $\phi =
\tan^{-1}(2\hbar\Omega/\alpha X^2)$].  This
will certainly be negligible if $\alpha X^2 T^*/2\hbar^2 \gg 1$, but we
cannot push $\alpha X^2/2\hbar$ down to the averaged level
spacing. Indeed, our deduction of (\ref{eq:final}) does not hold at
this level either, since there one should not neglect the upper limit
of (\ref{eq:upper}).

These conclusions are maintained if we substitute $N_{T^*}$ by the
integral of the smoothed density of section \ref{sec:III} in the
requantization of Bogomolny and Keating. The steps in the calculation
of the correlation function remain the same, so that now
$K^{k=0}(\Omega,X)$ is exactly (\ref{Kstandard}) with a finite
$\eta$. The expression for
\begin{eqnarray}
\ln \Lambda(\Omega_1,\Omega_2,X_1,X_2) & = & 2g\ \mbox{Re} \int
_\tau^\infty dt \frac{1}{t} \exp \left\{ i(\Omega_1 - \Omega_2)t/\hbar
- \left[ \alpha (X_1 - X_2)^2 /2\hbar^2 + 2\eta/\hbar \right] t
\right\} \nonumber \\ & = & 2g\ \mbox{Re}\ \mbox{E}_1 \left[
-i(\Omega_1 - \Omega_2)\tau/\hbar + \alpha (X_1 - X_2)^2\tau/2\hbar^2
+ 2\eta\tau/\hbar \right]
\end{eqnarray}
now holds, in principle, for any value of $\Omega_j$ or $X_j$,
including zero. We can now use the same approximation for the
exponential integral function as in (\ref{lnLambda}) to obtain
\begin{equation}
\Phi_k(\Omega_1,\Omega_2,X_1,X_2) = \left| \left[ -\frac{i}{\hbar}
(\Omega_1 - \Omega_2) + \frac{\alpha (X_1 - X_2)^2}{2\hbar^2} +
2\eta/\hbar \right] \frac{\hbar}{2\eta} \right|^{-2gk^2},
\end{equation}
which leads to (\ref{eq:final}) when we choose $\eta=\Delta/2\pi$ {\it and}
assume $\Omega$ and $\alpha X^2/2\hbar$ to be much greater than the
mean level spacing. In this sense, the energy smoothing with $\eta$
corresponds to the cutoff time $T^*$ ($\eta T^* \sim h$).

\section{Final discussion and conclusions}

The Gutzwiller series does not converge in the limit of small
smoothing. Indeed, it can be argued that the exponential smoothing
used in Section \ref{sec:III} is insufficient, but Gaussian smoothing
does work and leads to equivalent results. The application of the
periodic orbit theory for nongeneric long-range oscillations of the
energy level correlation functions was pushed in the last decade to a
range where contact could be made with the universal regime of Random
Matrix Theory. This involves the uniformity principle for long
periodic orbits, as well as the diagonal approximation. In our study
we have used the semiclassical trace formula to show the universality
of the level density parametric correlation function for classically
chaotic systems. This universality manifests itself after a proper
rescaling of the correlation function variables. Within the range of
validity of the semiclassical formulation and in the diagonal
approximation, we have obtained the same functional dependence on
energy and external parameter found by the method of RMT.

From a purely utilitary point of view, it may appear unnecessary to
rederive RMT results within a semiclassical theory. Yet, it is only in
this way that we can calculate the arbitrary parameters in RMT, as
well as derive the nongeneric long-range oscillations characteristic
of the individual systems that we wish to measure. In the present
case, we obtain the result that the unfolded mean square parametric
velocity diminishes by a factor of $\sqrt{2}$ when the time-reversal
symmetry is broken.

The semiclassical description of parametric correlations relies on the
ansatz proposed in (\ref{eq:gaussantz}), which is based on the central
limit theorem and leads to (\ref{Q2t}). To demonstrate rigorously the
validity of (\ref{eq:gaussantz}) for a generic classical chaotic
system, a systematic study of higher moments of $Q_\gamma$ is
required, which is a quite difficult task. Notwithstanding, there is
solid numerical evidence \cite{Bruus96,Pluhar95} to support the
proposed Gaussian ansatz. As a consequence of (\ref{Q2t}), $\alpha
X^2/2\hbar$ will always appear in the same footing as $\Omega$ in the
semiclassical approach. In general, this is not the case in the RMT,
as one can see in Section \ref{sec:II}. Only after linearizing the
``effective action" (what we do to generate the asymptotic expansion),
we see the semiclassical structure emerging.

It seemed that one would have to proceed beyond the diagonal
approximation to recover the oscillatory behavior which RMT predicts
at short scales. However, the remarkable requantization scheme
advanced by Bogomolny and Keating shows that it is only necessary to
feed in the discreteness of the quantum spectrum to obtain an
expression for the correlation functions that extrapolate to the
correct oscillatory RMT result in the asymptotic limit of large
$\Omega^2 + \alpha^2 X^4/4\hbar^2$. It is also important to stress
that only diagonal information about periodic orbits enters into this
result. In other words, nothing is said about correlations among orbit
actions, so that these may only affect higher terms in the asymptotic
expansion. In fact, to obtain higher-order corrections matching the
exact series, one would certainly need to introduce information about
inter-level correlations as well, which is apparently beyond the
capability of any present semiclassical approach, specially in the
case of arbitrary parameter variations.

The intrinsic limitation of the semiclassical method to cover small
energies at the quantum scale is remarkably manifest in the parametric
correlations. The lack of accuracy in energy ranges $\Omega \lesssim
\Delta$ imposes a limitation in the accuracy of parametric correlation
function for $X \lesssim X_c$, as one can see from
Eq.~(\ref{Kstandard}). In the standard derivation of
Sec. \ref{sec:III}, this is a direct consequence of the necessity of
smoothing the level density. Even in the requantization scheme,
although not explicitly, the same problem occurs, since both sharp and
smooth cutoffs in the Gutzwiller series do not assure convergence to
the actual eigenvalues.

Finally, remark that in this work we did not attempt to investigate
deviations from the universal, RMT behavior due to large scale
structures in the spectrum which can be ultimately related to short
periodic orbits of the system (see, for instance, the supersymmetric
treatment used in Ref. \onlinecite{Andreev95} for finite conductance
disordered systems). Again, the requantization scheme seems to be a
good starting point for such systematic studies from a purely
semiclassical point-of-view. Unfortunately, at present we only know
how to proceed safely by restricting ourselves to the energy range
where the diagonal approximation is accurate. There is a possibility
that the resummation technique could, in principle, extract additional
information from the $\Lambda$ function, but this work has still to be
done for generic systems. This is one of the major challenges of the
semiclassical theory.

\acknowledgments

The authors gratefully acknowledge Conselho Nacional de
Desenvolvimento Cient\'\i fico e Tecnol\'ogico (CNPq) for partial
financial support.

\appendix

\section{RMT asymptotic expansions}
\label{appendA}

Here we carry out the asymptotic expansion of $k(\omega,x)$ to lowest
order in $x^{-2}$ and $\omega^{-2}$ starting from the exact RMT
expressions (\ref{eq:k_GUE}), (\ref{eq:F_GUE}), (\ref{eq:k_GOE}), and
(\ref{eq:F_GOE}). Although, for the GUE, one could start from
(\ref{eq:k_GUEex}), we decided to use the integral expression to
illustrate the method. Essentially, we have to separate the
contributions coming from the two maximum-amplitude points. For the
GUE case, we begin by expanding $F^{GUE}$ around
$\lambda=\lambda_1=1$:
\begin{eqnarray}
k_{1,1}^{GUE}(\omega,x) & \approx & \frac{1}{2} \mbox{Re}
\int_0^\infty ds \int_0^\infty ds_1 \exp [ -(\pi^2 x^2 -
i\pi\omega^+)(s+s_1) ] \nonumber \\ & \approx & \frac{1}{2} \mbox{Re}
\left[ \frac{1}{(\pi^2 x^2 - i\pi\omega^+)^2} \right],
\label{eq:k11}
\end{eqnarray}
where $s \equiv 1 - \lambda$ and $s_1 \equiv \lambda_1 - 1$. Notice
that the integrand in (\ref{eq:k11}) is only appreciable for
$s,s_1\sim O(x^{-2})$ and, since we only need the lowest order in
$x^{-2}$, we can neglect quadratic terms in $s$. Equation
(\ref{eq:k11}), however, is still correct even if $x=0$, provided that
keep $\omega\gg 1$.

Next, we expand $F^{GUE}$ around $\lambda=-\lambda_1=-1$:
\begin{eqnarray}
k_{-1,1}^{GUE}(\omega,x) & \approx & \frac{1}{2} \mbox{Re}\
e^{2i\pi\omega^+} \int_0^\infty dr \int_0^\infty ds_1 \exp [ -\pi^2
x^2(r+s_1) - i\pi\omega^+(r-s_1) ] \nonumber \\ & \approx &
\frac{1}{2} \mbox{Re} \left( \frac{e^{2i\pi\omega^+}}{|\pi^2 x^2 -
i\pi\omega^+|^2} \right),
\label{eq:k_11}
\end{eqnarray}
where $r \equiv \lambda + 1$. This result is also valid for $x=0$ and
$\omega\gg 1$. Notice moreover that, contrary to the previous
contribution, this one is oscillating in $\omega$ on a scale $O(1)$
and therefore contains information about the discreteness of the
spectrum. We can add (\ref{eq:k11}) to (\ref{eq:k_11}) to obtain
Eq. (\ref{asympdgue}).

The GOE case is similar, but slightly more complicated due to the
additional structure in the integrand. Beginning with
$\lambda=\lambda_1=\lambda_2=1$, we can expand $F^{GOE}$ and the rest
of the integrand around this point to get
\begin{eqnarray}
k_{1,1,1}^{GOE}(\omega,x) & \approx & \mbox{Re} \int_0^\infty\! ds
\int_0^\infty\! ds_1 \int_0^\infty\! ds_2 \frac{(2s)(s+s_1+s_2)^2}
{(2ss_1+2ss_2-2s_1s_2+s_1^2+s_2^2+s^2)^2} 
\nonumber \\ & & \times \exp[
-(\pi^2 x^2/2 - i\pi\omega^+)(s+s_1+s_2)]  ~.
\label{eq:k111}
\end{eqnarray}
where $s_j = \lambda_j - 1$.
This integral can be evaluated by using spherical coordinates in
the first quadrant, giving
\begin{eqnarray}
k_{1,1,1}^{GOE}(\omega,x) & \!\approx \!& \mbox{Re} \left[
\frac{1}{(\pi^2 x^2/2 - i\pi\omega^+)^2} \right] \nonumber\\ && ~~~~\times
\int_0^{\pi/2}\!d\phi \int_0^{\pi/2} \!d\theta \frac{\sin 2\theta}{[1 +
\sin 2\theta (\cos\phi + \sin\phi) + (\cos 2\theta +
1)\sin\phi\cos\phi]^2}
\nonumber \\
&\approx & \mbox{Re} \left[
\frac{1}{(\pi^2 x^2/2 - i\pi\omega^+)^2} \right] ~.
\end{eqnarray}
(Here we have resorted to a numerical method to solve the double
integral.) Finally, we can expand around the other maximal point,
which is $\lambda=-\lambda_1=-\lambda_2=-1$, and obtain
\begin{eqnarray}
k_{-1,1,1}^{GOE}(\omega,x) & \approx & \mbox{Re}\ \Bigg\{
e^{2i\pi\omega^+} \int_0^\infty \!dr \int_0^\infty \!ds_1
\int_0^\infty \!ds_2 \frac{(2r)(-2)^2} {(-4)^2} \nonumber \\ & &
~~~~~\times \exp[ -(\pi^2 x^2/2)(r+s_1+s_2) - i\pi\omega^+(r-s_1-s_2)]
\Bigg\} \nonumber \\ & \approx & \frac{1}{2} \mbox{Re} \left(
\frac{e^{2i\pi\omega^+}}{|\pi^2 x^2/2 - i\pi\omega^+|^4} \right).
\label{eq:k_111}
\end{eqnarray}
Since this contribution is of higher order than (\ref{eq:k111}), we
should only keep the oscillating part and neglect the rest. With this
in mind, we arrive at Eq. (\ref{asympdgoe}).

\section{Parametric variation of periodic orbit}
\label{appendB}

It may be surprising that we can integrate the parametric velocity
$Q_\gamma$ in (\ref{defQ}), obtained from classical perturbation
theory to yield the exact result:
\begin{equation}
\label{daction}
S(E,X_2) - S(E,X_1)= \int_{X_1}^{X_2}\!dX \int_0^{T(E,X)}\! dt\,
\frac{\partial H}{\partial X} \Big(\xi(t), X \Big)
\end{equation}
where $\xi=({\bf p}, {\bf q})$ are vectors in the classical phase
space. Thus there is no limit on the period of the orbit if we keep
the right order of integration. We can obtain the expression above by
embedding the one parameter family of Hamiltonians $H(\xi, X)$ into a
single Hamiltonian ${\cal{H}}$ in a phase space that is expanded by
two more coordinates.  Hence, we add the parameter $X$ itself and a
conjugate variable $Y$, defining the Hamiltonian so that
\begin{equation}
{\cal{H}} (\xi, X, Y) \equiv H(\xi, X)
\end{equation}
at each point. Then, Hamilton's equations determine that $X$ is a
constant of the motion, whereas the equations for $\xi$ are unaltered,
so that the energy of the original systems is still constant, equal to
$E$. However, the periodic orbits of the original systems now
correspond to helicoidal orbits such that
\begin{equation}
\Delta Y = Q_\gamma(E, X) = \int_0^{T_\gamma} \! dt \,\dot{Y} (
\xi(t), X) = \int_0^{T_\gamma} \! dt\, \frac{\partial H}{\partial X}
\Big( \xi(t), X \Big) ~.
\end{equation}
Conservation of $E$ and $X$ implies that each of these helicoidal
orbits lies within a two-parameter family in the extended phase space.
Fixing the energy $E$, we thus obtain a two dimensional surface, along
which
\begin{equation}
\oint d{\bf q} \cdot {\bf p}   + \oint dX ~ Y - \oint dt~ {\cal{H}}   = 0
\end{equation}
for any reducible circuit (as a consequence of the Poincar\'e-Cartan
theorem \cite{Arnold}). The last integral cancels, because the full
energy has been chosen identical to the energies of the original
system, a constant along the surface. Picking the circuit so as to
connect two helicoids (corresponding to periodic orbits) with
different parameters, we obtain
\begin{equation}
\oint_{X_2} d{\bf q} \cdot {\bf p}   - 
\oint_{X_1} d{\bf q} \cdot {\bf p} = \int_{X_1}^{X_2} dX ~ Y ~,
\end{equation}
recovering (\ref{daction}).

It is important to note that the action difference refers to a
continuous family of periodic orbits. In the case where there is a set
of multiply symmetry-connected orbits, (\ref{daction}) can also be
used if the change of parameter does not break the symmetry.

We also point out that (\ref{daction}) refers strictly to an action
difference at constant energy. Of course, it is possible to choose
different one-parameter families of periodic orbits within the
two-parameter family in our problem. Goldberg {\it et al.}
\cite{Goldberg91} choose constant volume for the shell instead. This
coincides with constant energy for unfolded dynamical systems.

\section{Justification for the Gaussian ansatz}
\label{AppendC}

In this appendix we motivate the Gaussian ansatz used in
Ref.~\onlinecite{Bogomolny96} using arguments based solely on the
diagonal approximation. Let us start by Taylor expanding the
exponential term of $\Phi_k$ of Eq.~(\ref{phik}), thus recasting the
original exponential averaging problem into an averaging of increasing
powers of $(\delta N^{fl}_{T^*})^n$, with $\delta N^{fl}_{T^*}\equiv
N^{fl}_{T^*}(E, X) - N^{fl}_{T^*}(E^\prime, X^\prime)$.

The lowest-order term of the Taylor expansion ($n=1$), taken within
the validity range of classical perturbation theory, can be written as
\begin{eqnarray}
\Big\langle N^{fl}_{T^*}&&(E + \Omega_1, \overline{X} + X_1) -
N^{fl}_{T^*} (E + \Omega_2, \overline{X} + X_2) \Big\rangle_{E,
\overline{X}}= \nonumber\\ && \frac{1}{\pi^2}\left\langle \sum_{r,
T_\gamma<T^*} \frac{A_{\gamma r}} {r T_{\gamma}} \left[\exp\frac{i
rS_\gamma(E + \Omega_{1}, \overline{X} + X_1)}{\hbar} - \exp\frac{i
rS_\gamma(E + \Omega_{2}, \overline{X} + X_2)}{\hbar} \right]
\right\rangle_{E, \overline{X}},
\label{dNn1}
\end{eqnarray}
which clearly vanishes, since
\begin{equation}
\left\langle \mbox{e}^{i r S_\gamma (E, \overline{X})/\hbar} \left[
  \mbox{e}^{i r T_\gamma (E, \overline{X})\Omega_1/\hbar + i r
  Q_\gamma (E, \overline{X}) X_1/ \hbar} - \mbox{e}^{i r T_\gamma (E,
  \overline{X})\Omega_2/\hbar + i r Q_\gamma (E, \overline{X}) X_2/
  \hbar} \right] \right\rangle_{E, \overline{X}} = 0
\end{equation}
due to the very rapid oscillations of the term in $S_\gamma/\hbar$ in
the averaging interval $\delta E$.

Generally, in order to energy average the $n$-th power of $\delta
N^{fl}_{T^*}$ we have to calculate terms like
\begin{equation}
I_{\{r,\gamma\}}^{(n)} = \left\langle \exp\left[ \frac{1}{\hbar}(r_{1}
S_{\gamma_1} + r_{2} S_{\gamma_2} + \cdots + r_{n} S_{\gamma_n})
\right] \right\rangle_E ~.
\end{equation}
In fact, neglecting repetitions (see comment bellow), we conclude that
$I^{(n)} \approx 0$, whenever $n$ is odd, as in
(\ref{dNn1}). Moreover, for even values of $n$, we can always find a
set of actions that cancel each other, yielding
$I_{\{r,\gamma\}}^{(n)}\ne 0$. Due to phase-space restrictions, the
largest contribution will come from sets where the actions are grouped
pairwise. This allows $\langle(\delta N^{fl}_{T^*})^n \rangle$ to be
factorized into powers of $\langle(\delta N^{fl}_{T^*})^2\rangle$,
leading to the Gaussian formula used in
Ref.~\onlinecite{Bogomolny96}. Note, however, that our argument relies
on
\begin{equation}
\frac{1}{\hbar}(r_{1} S_{\gamma_1} + r_{2} S_{\gamma_2} + \cdots +
r_{n} S_{\gamma_n}) \gg 1 ~,
\label{eq:condC}
\end{equation}
for any distinct combination of trajectories. This condition is not
necessarily satisfied when $n \gg 1$. 

Finally, in a first inspection, the inclusion of repetitions would
seem to spoil our arguments in favour of the Gaussian ansatz. However,
using again phase-space considerations, it is simple to see that the
number of terms obtained by cancelling actions using repetitions is
much smaller than the number of simple pairwise cancellations of
primitive orbits. Thus the effect of the repetitions is not expected
to be important.


\end{document}